\documentclass[prb,twocolumn]{revtex4}

\usepackage{amsmath}

\begin{document}

\title{Constructing Fresnel reflection coefficients 
by ruler and compass}

\author{Juan J. Monz\'on and Luis L. S\'anchez-Soto}

\affiliation{Departamento de \'{O}ptica, 
Facultad de Ciencias F\'{\i}sicas, 
Universidad Complutense, 28040 Madrid, Spain}

\date{\today}

\begin{abstract}
A simple and intuitive geometical method to analyze
Fresnel formulas is presented. It applies to transparent
media and  is valid for perpendicular and parallel 
polarizations.  The approach gives a graphical 
characterization particularly simple of the critical 
and Brewster angles. It also provides an 
interpretation of the relation between the reflection 
coefficients for both basic polarizations as a symmetry 
in the plane.
\end{abstract}

\maketitle

\section{Introduction}

The reflection of a plane wave at a planar 
interface between two homogeneous and 
isotropic media is a well-known phenomenon.
From a general perspective, that embraces  
all  kinds of waves, the physics of reflection 
is well  understood: mismatched impedances 
generate the  reflected and transmitted waves, 
while the application of the proper boundary 
conditions at the discontinuity provides the 
corresponding amplitude coefficients~\cite{CR68}. 

For light waves the impedance is proportional to 
the refractive index.  Therefore, the mismatching
of impedances gives Snell's law, which, in fact,
is independent of the precise form of boundary
conditions. On the other hand, the continuity across 
the boundary of the tangential components of the 
electric and magnetic fields yields the reflection 
and transmission amplitudes, which constitute the 
famous Fresnel formulas~\cite{BO99,HE99,PE87}. 

In view of their simplicity and elegance, it seems 
difficult to say anything new about Fresnel formulas. 
However, a quick look at the index of, e.g., this 
journal~\cite{HE73,OL79,PA83,DO85,BE87,ZI88,NA89,LO91,DO92,RE92,MI96,OU01},
immediately reveals a steady flow of papers devoted 
to subtle aspects of this problem,  which shows that 
the  topic is far richer than one might naively expect.

Fresnel formulas provide complete optical information 
about the interface. Although it is possible to fully
examine their physical implications by a purely 
algebraic analysis, the discussion is usually carried
out by using plots of the coefficients in terms of
the angle of incidence.  This is mainly due to the
belief that graphical results convey information 
more readily than algebraic formulas. In this spirit, it
has also been  proposed the use of geometrical 
methods~\cite{LE22,KO75,DO80} to analyze the problem.
In our opinion, these geometrical approaches are
still worth exploring in order to take full advantage
of them. Thus,  in this paper we propose a 
new and extremely simple method that allows
one to construct Fresnel formulas by ruler and compass.
Geometrical construction by ruler and compass is a 
fascinating problem since ancient times, and offers
the additional advantage for the students of easily 
visualizing the steps of any construction and how it 
varies for different inputs.

We emphasize that these methods do not offer any 
inherent advantage in terms of computational efficiency. 
Apart from their beauty, their benefit lies in  the 
possibility of gaining insights into the qualitative 
behavior of the Fresnel coefficients, which is important 
in  developing a physical feeling of these relevant 
equations. 

\section{Fresnel formulas}

Let two homogeneous isotropic semi-infinite 
media, described by complex refractive indices 
$N_0$ and $N_1$, be separated by a plane 
boundary. We assume an incident monochromatic, 
linearly  polarized plane wave from medium 0, which 
makes an angle $\theta_0$ with the normal to 
the interface and has amplitude $E^i$. The 
electric field  is either in the plane of incidence 
(denoted by subscript $\parallel$) or perpendicular 
to the plane of incidence (subscript $\perp$). 
This wave splits into a reflected wave $E^r$ 
in medium 0, and a transmitted wave $E^t$ 
in medium 1 that makes an angle $\theta_1$ 
with the normal. The angles of incidence 
$\theta_0$ and refraction $\theta_1$ are related 
by Snell's law:
\begin{equation}
N_0 \sin \theta _0 = N_1 \sin \theta _1.  
\end{equation}
If media 0 and 1 are transparent (so that $N_0$ 
and  $N_1$ are real numbers) and no total reflection 
occurs, the angles $\theta_0$ and  $\theta_1$ are 
also real and the above picture of  how a plane 
wave is reflected and refracted at the interface 
is simple. However, when either one or both 
media is absorbing, the angles  $\theta_0$ and  
$\theta_1$  become, in general, complex and 
the  discussion continues to hold only  formally, 
but the physical picture of the fields becomes 
complicated~\cite{ST63}.

The wave vectors of all waves lie in the plane of 
incidence and when the incident fields are $\perp$- 
or $\parallel$-polarized, all  plane waves excited by 
the incident ones have the same polarization~\cite{AZ87}. 
An  arbitrarily polarized incident wave can be resolved into 
its $\perp$  and $\parallel$ components, and each of them
can be treated separately.

By demanding that the tangential components 
of $\mathbf{E}$ and  $\mathbf{H}$ should be 
continuous across the boundary~\cite{BO99},
and assuming nonmagnetic media, the reflection
and transmission amplitudes are given by~\cite{LE87}
\begin{subequations}
\begin{equation}
\label{rs}
r_\perp  =  \frac{E^r_\perp}
{E^i_\perp} = 
\frac{N_0 \cos\theta_0 -N_1 \cos\theta_1}
{N_0 \cos\theta_0 +N_1 \cos \theta_1},    
\end{equation}
\begin{equation}
\label{rp}
r _\parallel  =   \frac{E^r_ \parallel }
{E^i_\parallel}   =  
\frac{N_1 \cos\theta_0 -N_0 \cos\theta_1}
{N_1 \cos\theta_0 +N_0 \cos\theta_1},   
\end{equation}
\begin{equation}
t _\perp   =   \frac{E^t_\perp}
{E^i_\perp} =
\frac{2N_0 \cos\theta_0}
{N_0 \cos\theta_0 +N_1 \cos\theta_1} ,  
\end{equation}
\begin{equation}
t_\parallel  =    \frac{E^t_\parallel }
{E^i_\parallel}   =
\frac{2N_0 \cos\theta_0}
{N_1 \cos\theta_0 +N_0 \cos\theta_1},  
\end{equation}
\end{subequations}
which are the Fresnel formulas. The physical
contents of these coefficients are discussed 
in any optics textbook. 

\section{Graphical construction for the reflection coefficients}

\subsection{General considerations}

The geometrical method we outline here deals only with 
the case when media 0 and 1 are transparent. The refractive 
indices are then real numbers that we denote by $n_0$ and 
$n_1$. Before going into details we wish to show how the 
algebraic properties of the Fresnel reflection coefficients 
can be intimately linked to some elementary geometrical 
properties of an isosceles trapezoid. 

The key point for our purposes is to note that the reflection 
coefficients can be written as
\begin{equation}
\label{r}
r =  \frac{a-b}{a+b} ,
\end{equation}
where $a$ and $b$ are positive real numbers whose 
explicit form depend on the polarization. To visualize 
the algebraic properties of this quotient~\cite{LI98}
we propose to use the geometrical construction depicted
in Fig.~1. Let us assume first (Fig.~1.a) 
that $a > b$: then $r >0$ and this coefficient can be 
inferred from   an isosceles trapezoid  with minor basis $a-b$ 
and major  basis $a+b$, since the quotient of such bases is 
precisely  $r$. When $a = b$ (Fig.~1.b) the trapezoid degenerates 
in an isosceles triangle and $r=0$. Finally, when $a < b$, 
the numerator in $r$ is negative. Since a segment cannot 
have a negative length, we represent this case by plotting 
the bow tie of Fig.~1.c and we have then that $r < 0$.

In the next section we shall illustrate how this simple
construction allows for a complete determination of
the reflection coefficients. Before we do that and to
relate this interpretation with other previous results, 
we wish to note that any equation of the form (\ref{r}) 
can always be  expressed as
\begin{equation}
r =  \frac{a-b}{a+b} =
\frac{\xi - 1/\xi}{\xi + 1/\xi} =
\tanh \zeta ,
\end{equation}
where 
\begin{equation}
\xi = \sqrt{\frac{a}{b}} =  \exp(\zeta) .
\end{equation}
In spite of its simplicity, we think that this is a 
remarkable formula~\cite{KH79,CO91,MO99}.
It states that any reflection coefficient can be 
always expressed as a hyperbolic tangent, just as
in special relativity the velocities are expressed in 
terms of the rapidity~\cite{VG93,MO00,MO01}.

\subsection{Perpendicular polarization}

The previous general reasoning suggests that
Fresnel formulas can be seen from a purely
geometrical viewpoint. To go one step further,
and to quantify these ideas, we plot two concentric 
circumferences of radii $n_0$ and $n_1$ centered 
at the origin $O$, as shown in Fig.~2 (we need to 
take care only of the first quadrant). We assume 
$n_0<n_1$,  which does not suppose any serious 
restriction. 

For the case of polarization $\perp$ that we are 
considering, we draw in the circumference $n_0$ the 
radius that forms an angle $\theta_0$ with the horizontal. 
We denote by $S_0$ the point where this radius intersects 
the circumference. Next, we draw a horizontal line from 
$S_0$ that intersects the circumference $n_1$ at the
point $S_1$. The radius $OS_1$ determines then the 
refraction angle $\theta_1$.  A quick look at this figure 
shows that the projections on the vertical axis of the radii 
$OS_0$ and  $OS_1$ are identical, in agreement with 
Snell's law. 

If the light is incident from medium 1 at angle $\theta_1$ 
we would obtain  first  $S_1$ and, in a completely analogous 
way, the  corresponding pair $S_0$ and $\theta_0$.

When the point $S_0$ runs the quadrant  $n_0$ 
(i.e., $\theta_0$ varies from 0 to $\pi/2$)  the point 
$S_1$ runs the quadrant $n_1$ in such a way that 
$\theta_1$ varies from 0 to the critical angle $\theta_c$
(obviously, total internal reflection occurs only when the
light is incident from the denser medium). This is a clear
way of picturing the critical angle. 

On the other hand, the projections on the horizontal axis 
of the radii $OS_0$ and $OS_1$ measure $a= n_0\cos\theta_0$ 
and $b= n_1\cos\theta_1$, respectively. The values of $a-b$
and $a+b$ have been marked in Fig.~3 as bold segments. 
Since $a < b$ for every $\theta_0$, $r_\perp$ is always negative
and the trapezoid is a bow tie, according to our previous discussion.   
For $\theta_0 =0$ the bow tie degenerates into two horizontal 
segments of  lengths $n_1-n_0$ and $n_1+n_0$, while for  
$\theta_0 = \pi/2$ it reduces to two identical triangles.

As the angle of incidence increases, the minor basis grows 
while the major basis decreases. Therefore,  the  
absolute  value of $r_\perp$ grows monotonically 
with the angle of incidence, up to its maximum value 
of 1  at grazing incidence.

Finally, the numerical value of $r_\perp$ can be 
determined  by a ruler. To this end, it suffices with 
constructing  the right-angled triangle having as legs 
the bases of the previous bow tie.  It is obvious that  $r_\perp$  
coincides with the height of a  similar  triangle 
of unity basis, as shown in Fig.~3.

\subsection{Parallel polarization}

The richer phenomenology of the  $\parallel$ polarization 
is clearly highlighted also by this graphical construction.
The method is essentially the same as that for $\perp$ 
polarization. Once we are given the incidence angle 
$\theta_0$, we construct the refraction angle $\theta_1$ 
much in the same way.
 
It is clear from Eqs.~(\ref{rs}) and (\ref{rp}) that 
$r_\parallel$ can be obtained from $r_\perp$ by 
exchanging the refractive indices. In consequence, 
this suggests, as shown in Fig.~4, that now the radius 
$OS_0$ must be extended until it intersects the 
circumference $n_1$ at  the point $P_0$. Similarly, 
we obtain the point $P_1$ as the intersection of 
the radius $OS_1$ with the circumference $n_0$. 
Obviously, the horizontal projections of $OP_0$ and 
$OP_1$ are $a= n_1 \cos \theta_0$ and $b= n_0 \cos \theta_1$, 
respectively and, as before, $a-b$ and $a+b$ have been 
marked in Fig.~4  as bold segments.  These segments 
are the bases of an isosceles trapezoid that appears shaded 
in the figure. The nonparallel  sides have a length $n_0$ 
and form an angle $\theta_1$  with the major basis.   

For $\theta_0 =0$ the  trapezoid degenerates into two 
horizontal segments  of  lengths $n_1-n_0$ and
$n_1+n_0$, as it happens for  $\perp$ polarization.  
This reproduces the well-known fact that at normal 
incidence there is no physical difference between both 
basic polarizations, except for a  sign.

As $\theta_0$ increases, the minor basis $P_1 P_0^\prime$
decreases, reaching the value 0 (so $r_\parallel=0$ also), 
which defines the Brewster angle. In this particular angle 
($\tan \theta_B = n_1/n_0$) the points $P_0$ and $P_1$ 
are on the same vertical and, therefore, the trapezoid becomes 
a triangle. We think that this provides a remarkable way
of visualizing this important angle (see Fig.~5). When the  
angle of incidence is further increased,  $r_\parallel$ 
becomes negative. This can be seen in the fact that the 
trapezoid  becomes a  bow tie.  Finally, at grazing incidence 
this  bow tie is made from two identical isosceles triangles 
and then  $r_\parallel=-1$, as for the $\perp$ polarization.
The numerical value of $r_\parallel$ can be determined by
a ruler by the same procedure as before.

Before finishing, we wish to emphasize that, concerning 
the behavior of the corresponding transmission coefficients,
this geometrical construction still applies, but its interpretation 
is more involved, because they do not behave like a 
hyperbolic function~\cite{MO00}.

\subsection{Relation between basic polarizations}

It is the purpose of this section to show how our 
approach  interprets the quotient~\cite{HE99}
\begin{equation}
\label{quot}
\frac{r_\perp}{r_\parallel} =
- \frac{\cos(\theta_0 - \theta_1)}
{\cos(\theta_0 + \theta_1)} 
\end{equation}
as a simple symmetry in the plane. Azzam~\cite{AZ79a,AZ79b}  
has noticed that this relation can be recast so as to show
a universal character (i.e, it is independent of the two 
media that define the interface), and it has proven to be of 
practical interest. 

Let us denote by $\{ AB \}$ the horizontal
projection of the segment $AB$. From our 
previous discussion, we can recast the reflection 
coefficients as 
\begin{subequations}
\begin{equation}
\label{prs}
r_\perp =  \frac{\{OS_0 - OS_1\}}
{\{OS_0 + OS_1\} }= \frac{S_1S_0}
{\{OS\} } , 
\end{equation}
\begin{equation}
\label{prp}
r _\parallel  =   \frac{\{OP_0 - OP_1\}}
{\{OP_0 + OP_1\} }= \frac{\{P_1P_0\}}{\{OP\} }  ,
\end{equation}
\end{subequations}
where we have written $OS_0 + OS_1=OS$
and $OP_0 + OP_1=OP$. It is easy to check
that $OS$ and $OP$ have the same length.

Next, we note that the four points considered 
until now, namely, $S_0$, $S_1$ and $P_0$, $P_1$, 
are the vertex of the sector of annulus shaded in Fig.~6. 
The triangles $OP_0P_1$ and $OS_0S_1$
are identical and can be obtained by a reflection
through the bisecting line $OB$ of the sector.
The two diagonals $S_0S_1$ and $P_0P_1$  
have the same length and form between them
an angle $\theta_0 + \theta_1$.  On the other hand, 
the bisecting line $OB$ forms with the horizontal axis 
an angle $(\theta_0 + \theta_1)/2$ and it is a symmetry 
axis with respect to the pairs of points $S_0$ and $P_1$, 
$P_0$ and $S_1$, and $P$ and $S$. That is, $OP$ and
$OS$ form the same angle $\alpha$ with $OB$, where
$\alpha$ can be computed to be
\begin{eqnarray}
\tan \alpha & = & \frac{(n_1-n_0) \sin[(\theta_0 
- \theta_1)/2]}{(n_1+ n_0) \cos [(\theta_0 
- \theta_1)/2]} \nonumber \\
& = & \frac{n_1 - n_0}
{n_1+ n_0} \tan [(\theta_0 
- \theta_1)/2] .
\end{eqnarray}

We thus conclude that the numerators in 
(\ref{prs}) and (\ref{prp}) are related by
\begin{equation}
\frac{S_1S_0}{ \{ P_1 P_0 \} }
= \frac{1}{\cos(\theta_0 + \theta_1) } ,
\end{equation}
while for the denominators it holds
\begin{equation}
\frac{ \{ OP \} }{\{ O S \} }
= \frac{\cos[(\theta_0 + \theta_1)/2 + \alpha]}
{\cos[(\theta_0 + \theta_1)/2 - \alpha]}
= \cos(\theta_0 - \theta_1) .
\end{equation}
This concludes the geometrical proof of 
Eq.~(\ref{quot}) and shows that the coefficients 
$r_\perp$ and  $r_\parallel$ can be interpreted 
as quotients of horizontal  projections of 
segments that are symmetric with respect to 
the bisecting line. As far as we know, this simple
result has been never noticed in the literature.

\section{Concluding remarks}

The method given in this paper provides a 
geometrical construction for getting the values
of the Fresnel reflection coefficients by ruler
and compass. These graphical methods 
appeal more readily to students than an 
analytical treatment.

The construction allows one to show at a 
simple glance the variation of these coefficients 
with the angle of incidence and the peculiarities 
of the parallel polarization when crossing the Brewster angle.

\begin{acknowledgments}
We wish to thank B. Rose, F. S\'anchez-Quesada and
M. Sancho for a careful reading of the manuscript.
\end{acknowledgments}

\clearpage

\begin{figure}
\caption{Geometrical interpretation of the
Fresnel reflection coefficients as the 
quotient of two segments: a) for $a> b$ 
we have an isosceles  trapezoid and 
$r > 0$; b) for $a=b$ the trapezoid becomes a 
triangle and $r=0$; c) for $a < b$ then $r < 0$
and, by continuity, we represent now the 
trapezoid as a bow tie.}
\end{figure}

\begin{figure}
\caption{Illustrating how to obtain the
angle of refraction $\theta_1$ from the
angle of incidence $\theta_0$. The
appearance of the critical angle $\theta_c$
is evident.}
\end{figure}

\begin{figure}
\caption{Sketch of the method for getting the 
reflection coefficient $r_\perp$ by ruler and
compass.}
\end{figure}

\begin{figure}
\caption{Construction of the trapezoid associated
to the reflection coefficient $r_\parallel$ below 
the Brewster angle.}
\end{figure}

\begin{figure}
\caption{Same as in Fig.~4, but for the Brewster
angle.}
\end{figure}

\begin{figure}
\caption{Showing the relation between the
reflection coefficients for both basic
polarizations as a symmetry in the plane.}
\end{figure}

\end{document}